\documentclass[12pt,a4paper]{article}
\usepackage[dvips]{graphicx}

\begin{document}
\begin{flushright}
UAB--FT--649\\
October 2008
\end{flushright}
\vspace*{0.6cm}

\begin{center}
{\Large\bf
Linear Sigma Model predictions for $V\to S\gamma$ and $S\to V\gamma$ decays}
\vspace*{1cm}

Rafel Escribano$^1$, Pere Masjuan$^1$ and Jordi Nadal$^2$ 
\vspace*{0.2cm}

{\footnotesize\it
$^1$Grup de F\'{\i}sica Te\`orica and IFAE, Universitat Aut\`onoma de Barcelona,\\
E-08193 Bellaterra (Barcelona), Spain

$^2$IFAE, Universitat Aut\`onoma de Barcelona, E-08193 Bellaterra (Barcelona), Spain}

\end{center}
\vspace*{0.2cm}

\begin{abstract}
The radiative decays $V\to S\gamma$ and $S\to V\gamma$ with
$V=\rho, \omega, \phi$ and $S=a_0, \sigma, f_0$ are calculated within the framework of the
Linear Sigma Model.
Current experimental data on the $\phi\to f_0\gamma$ and $\rho\to\sigma\gamma$
branching ratios and the ratio $\Gamma(\phi\to f_0\gamma)/\Gamma(\phi\to a_0\gamma)$
are satisfactorily accommodated in our approach.
We also estimate the decay widths of the $f_0,a_0\to\rho\gamma,\omega\gamma$ transitions.
All the processes considered are of interest for ongoing experimental programs in Frascati, J\"ulich and Novosibirsk.
\end{abstract}

\section{Introduction}
\label{intro}
The nature and properties of the lowest lying scalar states have been and still are a controversial issue
\cite{Close:2002zu,Amsler:2004ps,Amsler:2008zz}.
Concerning their properties, only the ones of the $I=1$ $a_0(980)$ and $I=0$ $f_0(980)$ scalar mesons are well established \cite{Amsler:2008zz}.
The $I=0$ $f_0(600)$ or $\sigma$ meson is now confirmed and its properties start to be known
while the $I=1/2$ $K^\ast_0(800)$ or $\kappa$ meson is still omitted from the summary table of the most recent version of the Review of Particle Physics \cite{Amsler:2008zz} and needs confirmation.
With respect to the nature, it is yet not clear whether they are $q\bar q$ states, four-quark states, meson-meson molecules, glueballs, or most likely a mixture of all of them.
The decay of scalar mesons into two photons has been proposed as a method to discriminate among different models regarding their nature.
For instance, the width of $\sigma\to\gamma\gamma$ has been obtained from the reaction
$\gamma\gamma\to\pi^0\pi^0$ \cite{Pennington:2006dg,Oller:2007sh} or, more recently, from the nucleon electromagnetic polarizabilities \cite{Bernabeu:2008wt}.
However, the obtained value of the $\sigma$ width depends upon the accurate description of the production amplitude and different parametrizations give somewhat distinct results.

Due to the complexity of the former analyses and the scarce experimental data available,
the radiative $V\to S\gamma$  decays have also been proposed, similarly to $S\to\gamma\gamma$ decays, as sensitive reactions to distinguish among different models about the nature of scalar mesons \cite{Amsler:2004ps}.
Unlike $S\to\gamma\gamma$ decays, the width of the various $V\to S\gamma$ decays can be extracted from a much more clean experimental environment through the analysis of precise data on
$V\to P^0P^0\gamma$ decays, via the decay chain $V\to S\gamma\to P^0P^0\gamma$, after background subtraction.
On the experimental side,
the KLOE Collaboration has reported the following measurements:
$B(\phi\to f_0\gamma)=(3.21^{+0.03}_{-0.09}\pm 0.18)\times 10^{-4}$ \cite{Ambrosino:2006hb} and
$B(\phi\to a_0\gamma)=(7.4\pm 0.7)\times 10^{-5}$ \cite{Aloisio:2002bsa},
from the analysis of $\phi\to\pi^0\pi^0\gamma$ and $\phi\to\pi^0\eta\gamma$ decays, respectively, and
$B(\phi\to f_0\gamma)/B(\phi\to a_0\gamma)=6.1\pm 0.6$ \cite{Aloisio:2002bsa},
which agree with the CMD-2 and SND Coll.~values
$B(\phi\to f_0\gamma)=(2.90\pm 0.21\pm 1.54)\times 10^{-4}$ \cite{Akhmetshin:1999di} and
$B(\phi\to a_0\gamma)=(8.8\pm 1.7)\times 10^{-5}$ \cite{Achasov:2000ku}.
Finally, the Particle Data Group (PDG) fit gives
$B(\phi\to f_0\gamma)=(3.22\pm 0.19)\times 10^{-4}$ and
$B(\phi\to a_0\gamma)=(7.6\pm 0.6)\times 10^{-5}$ \cite{Amsler:2008zz}.
The CMD-2 Coll.~has also reported
$B(\rho\to\sigma\gamma)=(6.0^{+3.3}_{-2.7}\pm 0.9)\times 10^{-5}$ from the measurement of the
$e^+e^-\to\pi^0\pi^0\gamma$ cross section in the c.~m.~energy range 600--970 MeV
\cite{Akhmetshin:2003rg}.
On the phenomenological side,
the branching ratios of $\phi\to f_0\gamma$ and $\phi\to a_0\gamma$,
discussed originally in Ref.~\cite{Achasov:1987ts},
have been recently obtained in Ref.~\cite{Black:2002ek} from a vector meson dominance model 
(more precisely, the $\phi\to a_0\gamma$ branching ratio is taken as an input value while the
$\phi\to f_0\gamma$ one and their ratio are predicted in two variants of the model),
in Ref.~\cite{Kalashnikova:2005zz} using quark or meson loops as the driven mechanism of the decays, and in Ref.~\cite{Ivashyn:2007yy} within a unified ChPT-based approach at one-loop level
(the $\phi\to f_0\gamma$ branching ratio is now taken as input).
See also the update of the previous work in Ref.~\cite{Ivashyn:2008sg}.
Besides the former measurements by the KLOE, CMD-2 and SND Coll., the reactions
$f_0\to\rho\gamma, \omega\gamma$ and $a_0\to\rho\gamma, \omega\gamma$
are currently under study at WASA@COSY
(see Ref.~\cite{Buscher:2008ge} and references therein).
Different predictions for these decays can also be found in
Refs.~\cite{Black:2002ek,Kalashnikova:2005zz,Ivashyn:2007yy,Ivashyn:2008sg} as well as
in Ref.~\cite{Nagahiro:2008mn} where they have been obtained using a Chiral Unitary Approach.
For the sake of comparison, the numerical values of all these results are shown in Table
\ref{table2}.
As seen, the predictions for these decays are affected by model details,
thus precise experimental measurements would be highly desirable to differentiate among models.

The aim of this Letter is to calculate all possible $V\to S\gamma$ and $S\to V\gamma$ decays,
with $V=\rho, \omega, \phi$ and $S=a_0, \sigma, f_0$, within the framework of the Linear Sigma Model (L$\sigma$M).
The advantage of using this model is that it satisfies all the chiral constraints and also incorporates the scalar mesons in an explicit way.
In particular, this model fixes the required scalar-pseudoscalar-pseudoscalar couplings in terms of scalar and pseudoscalar masses and pseudoscalar decay constants\footnote{
The masses of the $a_0$ and $\kappa$ mesons are fixed within the L$\sigma$M in terms of pseudoscalar masses and decay constants.
In contrast, the $f_0$ mass has to be taken from experiment in order to fix completely the $I=0$ scalar sector, \textit{i.e.}~the $\sigma$ mass and the mixing angle \cite{Napsuciale:1998ip}.
The resultant values for these parameters are close to the measured ones.}.
The drawback is that the strong amplitudes, being used at tree level, do not fulfill unitarity exactly
(unlike Unitarized ChPT where exact unitarity is implemented in coupled channels,
see for instance Ref.~\cite{Oset:2002zd} and references therein).
In any case, the L$\sigma$M has been shown to be very useful in predicting adequately the spectra and the integrated branching ratios of $V\to P^0P^0\gamma$ decays
\cite{Escribano:2006mb,Escribano:2006az}, where in most cases of interest the signal is dominated or contributed significantly by the intermediate mechanism $V\to S\gamma$.
In Sec.~\ref{VSgamma}, we calculate the radiative decays $V\to S\gamma$ and $S\to V\gamma$
in the L$\sigma$M and compare the obtained results with other approaches.
Finally, we conclude in Sec.~\ref{conclusions}.

\section{$V\to S\gamma$ and $S\to V\gamma$ decays}
\label{VSgamma}
The $V\to S\gamma$ decays can be treated in a L$\sigma$M extended to incorporate external vector mesons and photons.
This may be simply achieved by means of the Lagrangian
\begin{equation}
\label{LsMLag}
{\cal L}=\frac{1}{2}\langle D_\mu\Sigma^\dagger D^\mu\Sigma\rangle+
{\cal L}_{\rm SPP}+{\cal L}_{\rm SSS}+\cdots\ ,
\end{equation}
where $\Sigma=S+iP$, $S$ and $P$ are the pseudoscalar nonet matrices
and the covariant derivative is defined as
$D_\mu\Sigma=\partial_\mu\Sigma-i e A_\mu [Q,\Sigma]-i g [V_\mu,\Sigma]$ with
$Q=\mbox{diag}(2/3,-1/3,-1/3)$ being the quark charge matrix and
$V_\mu$ the additional matrix containing the nonet of vector mesons.
${\cal L}_{\rm SPP}$ and ${\cal L}_{\rm SSS}$ contain the required scalar-pseudoscalar-pseudoscalar and scalar-scalar-scalar couplings, respectively, while the dots stand for interactions not relevant to our analysis (see Ref.~\cite{Napsuciale:1998ip} for a detailed discussion of the different terms contributing to the L$\sigma$M Lagrangian).

\begin{figure}[t]
\centerline{\includegraphics[width=\textwidth]{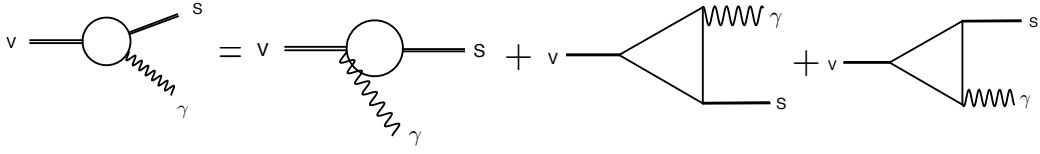}}  
\caption{\small
Feynman diagrams contributing to $V\to S\gamma$ processes.}
\label{diagVSgamma}
\end{figure}

One easily observes from the Lagrangian (\ref{LsMLag}) that there are no contact
$VS\gamma$ terms at the lowest (tree-level) order.
Therefore, the leading contribution includes the one-loop diagrams with intermediate pseudoscalar and scalar mesons shown in Fig.~\ref{diagVSgamma}.
A straightforward calculation of $V(q^\ast,\epsilon^\ast)\to S(p)\gamma(q,\epsilon)$ leads to a
\emph{finite} amplitude that is conveniently parametrized in the following way:
\begin{equation}
\label{AVSgamma}
{\setlength\arraycolsep{2pt}
\begin{array}{rcl}
{\cal A}^\pi_{\rho\to\sigma\gamma}&=&
\frac{-e g}{\sqrt{2}\pi^2 m_{\pi^+}^2}\{a\}
L(m_{\rho\pi}^2,m_{\sigma\pi}^2)g_{\sigma\pi^+\pi^-}\ ,\\[3ex]
{\cal A}^K_{(\rho,\omega)\to\sigma\gamma}&=&
\frac{-e g}{2\sqrt{2}\pi^2 m_{K^+}^2}\{a\}
L(m_{(\rho,\omega)K}^2,m_{\sigma K}^2)g_{\sigma K^+K^-}\ ,\\[3ex]
{\cal A}^K_{\phi\to(\sigma,f_0,a_0)\gamma}&=&
\frac{e g}{2\pi^2 m_{K^+}^2}\{a\}
L(m_{\phi K}^2,m_{(\sigma,f_0,a_0)K}^2)g_{(\sigma,f_0,a_0) K^+K^-}\ ,\\[3ex]
{\cal A}^\pi_{f_0\to\rho\gamma}&=&
\frac{e g}{\sqrt{2}\pi^2 m_{\pi^+}^2}\{a\}
L(m_{\rho\pi}^2,m_{f_0\pi}^2)g_{f_0\pi^+\pi^-}\ ,\\[3ex]
{\cal A}^K_{(f_0,a_0^0)\to(\rho^0,\omega)\gamma}&=&
\frac{e g}{2\sqrt{2}\pi^2 m_{K^+}^2}\{a\}
L(m_{(\rho,\omega)K}^2,m_{(f_0,a_0)K}^2)g_{(f_0,a_0) K^+K^-}\ ,\\[1ex]
\end{array}}
\end{equation}
where $\{a\}=(\epsilon^\ast\epsilon)(q^\ast q)-(\epsilon^\ast q)(\epsilon q^\ast)$
makes the amplitude Lorentz and gauge-invariant,
$L(m_{VP}^2,m_{SP}^2)$ is the loop integral function defined as \cite{Close:1992ay}
\begin{equation}
\label{L}
\begin{array}{rl}
L(a,b) &=\ 
\frac{1}{2(a-b)}-
\frac{2}{(a-b)^2}\left[f\left(\frac{1}{b}\right)-f\left(\frac{1}{a}\right)\right]+\\[2ex]
&+\ \frac{a}{(a-b)^2}\left[g\left(\frac{1}{b}\right)-g\left(\frac{1}{a}\right)\right]\ 
\end{array}
\end{equation}
with 
\begin{equation}
\label{f&g}
\begin{array}{l}
f(z)=\left\{
\begin{array}{ll}
-\left[\arcsin\left(\frac{1}{2\sqrt{z}}\right)\right]^2 & z>\frac{1}{4}\\[1ex]
\frac{1}{4}\left(\log\frac{\eta_+}{\eta_-}-i\pi\right)^2 & z<\frac{1}{4}
\end{array}
\right.\\[5ex]
g(z)=\left\{
\begin{array}{ll}
\sqrt{4z-1}\arcsin\left(\frac{1}{2\sqrt{z}}\right) & z>\frac{1}{4}\\[1ex]
\frac{1}{2}\sqrt{1-4z}\left(\log\frac{\eta_+}{\eta_-}-i\pi\right) & z<\frac{1}{4}
\end{array}
\right.
\end{array}
\end{equation}
and
$\eta_\pm=\frac{1}{2z}(1\pm\sqrt{1-4z})$,
$m_{VP}^2=m_V^2/m_{P^+}^2$ and $m_{SP}^2=m_S^2/m_{P^+}^2$.
The $SPP$ coupling constants are fixed within the L$\sigma$M to
\begin{equation}
\label{ASPP}
\begin{array}{c}
g_{\sigma\pi^+\pi^-}=\frac{m^2_\pi-m^2_\sigma}{f_\pi}\cos\phi_S\ ,
\quad
g_{f_0\pi^+\pi^-}=\frac{m^2_\pi-m^2_{f_0}}{f_\pi}\sin\phi_S\ ,\\[2ex]
g_{\sigma K^+K^-}=\frac{m^2_K-m^2_\sigma}{2f_K}(\cos\phi_S-\sqrt{2}\sin\phi_S)\ ,
\quad
g_{a_0 K^+K^-}=\frac{m^2_K-m^2_{a_0}}{2f_K}\ ,\\[2ex]
g_{f_0 K^+K^-}=\frac{m^2_K-m^2_{f_0}}{2f_K}(\sin\phi_S+\sqrt{2}\cos\phi_S)\ ,
\end{array}
\end{equation}
where $\phi_S$ is the scalar mixing angle in the quark flavour basis defined as
\begin{equation}
\begin{array}{c}
\sigma=\cos\phi_S\sigma_q-\sin\phi_S\sigma_s\ ,\\[2ex]
f_0=\sin\phi_S\sigma_q+\cos\phi_S\sigma_s\ ,
\end{array}
\end{equation}
with $\sigma_q\equiv\frac{1}{\sqrt{2}}(u\bar u+d\bar d)$ and $\sigma_s\equiv s\bar s$
\cite{Napsuciale:1998ip}.
The coupling constant $g$ comes from the strong amplitude 
${\cal A}(\rho\rightarrow\pi^+\pi^-)=-\sqrt{2}g\epsilon^\ast (p_+-p_-)$  
with $|g|\simeq 4.2$ to agree with $\Gamma(\rho\rightarrow\pi^+\pi^-)_{\rm exp}= 147.8$ MeV. 
However, for the $\phi$ decays we replace $g$ by $g_{s}$ where $|g_s|\simeq 4.5$
to agree with $\Gamma(\phi\rightarrow K^+K^-)_{\rm exp}= 2.10$ MeV \cite{Amsler:2008zz}
---in the good $SU(3)$ limit one should have $|g|=|g_{s}|$.
These couplings are the part beyond the L$\sigma$M which we have fixed phenomenologically.

In addition to the former amplitudes driven by pseudoscalar meson loops,
the appearance of $SSS$ couplings in the L$\sigma$M Lagrangian (\ref{LsMLag}) allows for scalar meson loops as well.
Accordingly, the following $\kappa$ induced amplitudes have also to be taken into consideration:
\begin{equation}
\label{AVSgammaS}
{\setlength\arraycolsep{2pt}
\begin{array}{rcl}
{\cal A}^\kappa_{(\rho,\omega)\to\sigma\gamma}&=&
\frac{-e g}{2\sqrt{2}\pi^2 m_{\kappa^+}^2}\{a\}
L(m_{(\rho,\omega)\kappa}^2,m_{\sigma\kappa}^2)g_{\sigma\kappa^+\kappa^-}\ ,\\[3ex]
{\cal A}^\kappa_{\phi\to(\sigma,f_0,a_0)\gamma}&=&
\frac{e g}{2\pi^2 m_{\kappa^+}^2}\{a\}
L(m_{\phi\kappa}^2,m_{(\sigma,f_0,a_0)\kappa}^2)g_{(\sigma,f_0,a_0)\kappa^+\kappa^-}\ ,\\[3ex]
{\cal A}^\kappa_{(f_0,a_0^0)\to(\rho^0,\omega)\gamma}&=&
\frac{e g}{2\sqrt{2}\pi^2 m_{\kappa^+}^2}\{a\}
L(m_{(\rho,\omega)\kappa}^2,m_{(f_0,a_0)\kappa}^2)g_{(f_0,a_0)\kappa^+\kappa^-}\ ,\\[1ex]
\end{array}}
\end{equation}
where $m_{V\kappa}^2=m_V^2/m_\kappa^2$ and $m_{S\kappa}^2=m_S^2/m_\kappa^2$.
Chiral symmetry requires $VPP$ and $VSS$ contact terms to have the same coupling constants.
For this reason, the amplitudes in Eq.~(\ref{AVSgammaS}) are also proportional to $g$ and formally identical to those in Eq.~(\ref{AVSgamma}) except for the replacements $m_K^2$ by $m_\kappa^2$ and $g_{SK^+K^-}$ by $g_{S\kappa^+\kappa^-}$.
The required $SSS$ couplings are \cite{Rodriguez:2004tn}
\begin{equation}
\label{ASSS}
\begin{array}{c}
g_{\sigma\kappa^+\kappa^-}=
-\frac{m^2_\kappa-m^2_\sigma}{2(f_K-f_\pi)}(\cos\phi_S+\sqrt{2}\sin\phi_S)\ ,
\quad
g_{a_0\kappa^+\kappa^-}=-\frac{m^2_\kappa-m^2_{a_0}}{2(f_K-f_\pi)}\ ,\\[2ex]
g_{f_0\kappa^+\kappa^-}=-\frac{m^2_\kappa-m^2_{f_0}}{2(f_K-f_\pi)}(\sin\phi_S-\sqrt{2}\cos\phi_S)\ .
\end{array}
\end{equation}

As seen from the amplitudes in Eqs.~(\ref{AVSgamma},\ref{AVSgammaS}),
due to $G$-parity and the Zweig rule, this latter for $\phi$ processes,
the $V\to S\gamma$ decays proceed trough charged kaon and kappa loops except for $\rho$ processes where charged pion loops are also allowed.
Squaring these amplitudes and averaging over polarizations for the vector meson initiated processes one gets the kaon, pion and kappa loop contributions to the different
$V\to S\gamma$ and $S\to V\gamma$ decays shown in Table \ref{table1}.
The total contribution, namely the sum of the former contributions plus their respective interferences, is also shown for each decay.
Our results have been obtained using the following numerical input:
$f_\pi=92.3$ MeV and $f_K=1.196 f_\pi$ from the recent review \cite{Rosner:2008yu},
$m_\sigma=513\pm 32$ MeV \cite{Muramatsu:2002jp},
$m_{a_0}=984.7\pm 1.2$ MeV \cite{Amsler:2008zz},
$m_{f_0}=985$ MeV and $\phi_S=-8^\circ$ from the best fit to $\phi\to\pi^0\pi^0\gamma$ data performed in Ref.~\cite{Escribano:2006mb}.
For the rest of parameters we use the values reported in Ref.~\cite{Amsler:2008zz}.
The error associated to the kappa loop contributions and the total error (if relevant) come from the uncertainty, attributed by the PDG \cite{Amsler:2008zz}, about the kappa mass,
$m_\kappa=672\pm 40$ MeV.
These kappa loop contributions turn out to be insignificant for $\phi\to f_0\gamma,a_0\gamma$ decays  and dominant for $\omega,\phi\to\sigma\gamma$.
This behaviour is  easily understood in terms of the $\sigma K\bar K$ coupling which in the L$\sigma$M is proportional to the difference $m_\sigma^2-m_K^2$ and thus negligible for $m_\sigma\simeq m_K$.
The same would happen to the $\rho\to\sigma\gamma$ decay but in this case the allowed pion loop contribution plays the dominant role.
Regarding the $f_0,a_0\to\rho\gamma,\omega\gamma$ decays,
the interference of the kappa loop contributions, not important in absolute magnitude,
with the pion and/or kaon ones slightly modify the results (15\% or less).
In contrast, the interference between pion and kaon loop contributions in the
$f_0\to\rho\gamma$ case alters the final value in a substantial amount (40\% approx.).
\begin{table}
\centerline{
\begin{tabular}{lccccc}
\hline\\[-2ex]
Observable &
$K$ &
$\pi$ &
$\kappa$ &
Total\\[0.5ex]
\hline\\[-2ex]
$B_{\phi\to f_0\gamma}\ [10^{-4}]$ &
$2.3$ & --- & $(8.4^{+4.6}_{-3.1})\times 10^{-4}$ & $2.3$\\[0.5ex]
$B_{\phi\to a_0\gamma}\ [10^{-4}]$ &
$1.5$ & --- & $(3.6^{+2.0}_{-1.3})\times 10^{-4}$ & $1.5$\\[0.5ex]
$B_{\phi\to\sigma\gamma}\ [10^{-5}]$ &
$0.20$ & --- & $2.5^{+0.9}_{-0.8}$ & $1.4\pm 0.6$\\[0.5ex]
$B_{\rho\to\sigma\gamma}\ [10^{-5}]$ &
$2.1\times 10^{-4}$ & $9.9$ & $(1.1^{+0.3}_{-0.4})\times 10^{-2}$ & $10.5^{+0.1}_{-0.2}$\\[0.5ex]
$B_{\omega\to\sigma\gamma}\ [10^{-5}]$ & 
$4.0\times 10^{-3}$ & --- & $0.20^{+0.06}_{-0.07}$ & $0.15^{+0.05}_{-0.06}$\\[0.5ex]
\hline\\[-2ex]
$\Gamma_{f_0\to\rho\gamma}$ [keV] &
$17.6$ & $1.1$ & $0.12^{+0.06}_{-0.05}$ & $24.1^{+0.7}_{-0.8}$\\[0.5ex]
$\Gamma_{f_0\to\omega\gamma}$ [keV] &
$16.6$ & --- & $0.10^{+0.06}_{-0.03}$ & $14.1^{+0.4}_{-0.6}$\\[0.5ex]
$\Gamma_{a_0^0\to\rho^0\gamma}$ [keV] &
$10.9$ & --- & $0.05^{+0.03}_{-0.02}$ & $12.4\pm 0.3$\\[0.5ex]
$\Gamma_{a_0\to\omega\gamma}$ [keV] &
$10.2$ & --- & $0.04^{+0.03}_{-0.01}$ & $11.6\pm 0.3$\\[0.5ex]
\hline
\end{tabular}
}
\caption{
L$\sigma$M contributions from $K$, $\pi$ and $\kappa$ loops to $V\to S\gamma$ and $S\to V\gamma$ decays.}
\label{table1}
\end{table}

In Tables \ref{table2} and \ref{table3}, our final values together with the results of several recent approaches as well as the reported experimental measurements are shown for comparison.
We start discussing the $\phi\to f_0\gamma$ process.
Our central value is in fair agreement with the experimental measurement.
However, the predicted branching ratio is very sensitive to the $f_0$ mass due to the steep behaviour of the kaon loop function after threshold ($2m_K\simeq 991$ MeV).
In addition, the value of the loop function itself changes if the kaon threshold is taken to be the charged one ($2m_{K^+}\simeq 987$ MeV).
For such reasons, we prefer to be conservative and give errors to our predictions which include all possible values of the branching ratios ranging from $m_{f_0}=970$ MeV to 990 MeV
and are independent of the threshold chosen.
For the $\phi\to a_0\gamma$ process, our predicted value is compatible with other approaches but disagrees with the measured branching ratio.
The value found would decrease for $a_0$ masses higher than the kaon threshold.
For instance, the experimental value is obtained for $m_{a_0}\simeq 992$ MeV.
In this case, however, the $a_0$ mass is very well measured and the error of our prediction is smaller.
\begin{table}
\centerline{\footnotesize
\begin{tabular}{lcccccc}
\hline\\[-2ex]
Observable &
L$\sigma$M &
M1 \cite{Black:2002ek} &
M2 \cite{Black:2002ek} &
\cite{Kalashnikova:2005zz} &
\cite{Ivashyn:2007yy} &
PDG \cite{Amsler:2008zz}\\[0.5ex]
\hline\\[-2ex]
$B_{\phi\to f_0\gamma}\ [10^{-4}]$ &
$2.3^{+1.0}_{-1.2}$ &
$0.49\pm 0.07$ &
$0.49\pm 0.07$ &
$1.4$ &
--- &
$3.22\pm 0.19$\\[0.5ex]
$B_{\phi\to a_0\gamma}\ [10^{-5}]$ &
$14.8^{+2.7}_{-0.4}$ &
--- &
--- &
$14$ &
$16.7$ &
$7.6\pm0.6$\\[0.5ex]
$R_{\phi\to f_0\gamma/a_0\gamma}$ &
$1.5^{+0.4}_{-0.9}$ &
$0.26\pm 0.06$ &
$0.46\pm 0.09$ &
$1$ &
$2.6$ &
$6.1\pm 0.6$\\[0.5ex]
\hline\\[-2ex]
$B_{\phi\to\sigma\gamma}\ [10^{-3}]$ &
$[0.2,6.8]\times 10^{-2}$ &
$32.2\pm 4.5$ &
$7.7\pm 2.1$ &
--- &
--- &
---\\[0.5ex]
$B_{\rho\to\sigma\gamma}\ [10^{-5}]$ &
$10.5^{+2.9}_{-2.5}$ &
$0.15\pm 0.31$ &
$11\pm 3$ &
--- &
--- &
$6.0^{+3.3}_{-2.7}\pm 0.9$\\[0.5ex]
$B_{\omega\to\sigma\gamma}\ [10^{-3}]$ &
$[0.2,5.4]\times 10^{-3}$ &
$1.9\pm 0.4$ &
$3.9\pm 0.5$ &
--- &
--- &
---\\[0.5ex]
\hline
\end{tabular}
}
\caption{
Comparison of L$\sigma$M predictions for $V\to S\gamma$ decays with other approaches and experimental measurements if available.}
\label{table2}
\end{table}
The ratio of the two previous branching ratios is given by
\begin{equation}
\label{Rf0a0}
R_{\phi\rightarrow f_{0}\gamma/a_{0}\gamma}^{\mbox{\scriptsize L$\sigma$M}}=
\frac{|L(m^2_{f_{0}})|^2}{|L(m^2_{a_{0}})|^2}
\frac{\left(1-m^2_{f_{0}}/m^2_{\phi}\right)^3}
       {\left(1-m^2_{a_{0}}/m^2_{\phi}\right)^3}
       \times\frac{g^2_{f_{0}K^+K^-}}{g^2_{a_{0}K^+K^-}}
\simeq ({\rm s}\phi_S +\sqrt{2}{\rm c}\phi_S)^2\ ,
\end{equation}
where the approximation is valid for $m_{f_{0}}\simeq m_{a_{0}}$ with both masses above or below threshold.
Again, our prediction is consistent with other analyses but differs from the experimental measurement.
Nevertheless, the PDG reported value is obtained from
$B(\phi\to f_0\gamma)=(4.47\pm 0.21)\times 10^{-4}$, which is 40\% bigger than the current fitted value \cite{Amsler:2008zz}.
This large branching ratio was found in Ref.~\cite{Aloisio:2002bt}
due to a large destructive interference between the $f_0\gamma$ and $\sigma\gamma$ contributions to $\phi\to\pi^0\pi^0\gamma$, in disagreement with other experiments \cite{Akhmetshin:1999di}.
If instead, the ratio of the $\phi\to f_0\gamma$ and $\phi\to a_0\gamma$ branching ratios is calculated from their present-day values \cite{Amsler:2008zz} one gets
$R_{\phi\rightarrow f_{0}\gamma/a_{0}\gamma}=4.2\pm 0.4$, in better agreement with our prediction.
It is worth mentioning that in the three observables discussed so far, the uncertainty about the choice of kaon threshold is more important than the one caused by the $f_0$ and $a_0$ masses.
$B(\phi\to f_0\gamma)$ and $R_{\phi\rightarrow f_{0}\gamma/a_{0}\gamma}$ also depend on the scalar mixing angle.
For comparison, one gets $B(\phi\to f_0\gamma)=2.9\times 10^{-4}$ and 
$R_{\phi\rightarrow f_{0}\gamma/a_{0}\gamma}=1.9$ for $\phi_S=0^\circ$.
One should keep in mind, however, that our approach has been shown to be rather accurate in
predicting the spectra and the integrated branching ratios of $\phi\to\pi^0\pi^0\gamma$ and
$\phi\to\pi^0\eta\gamma$ decays, where the scalar contributions
driven by $\phi\to f_0\gamma$ and $\phi\to a_0\gamma$ respectively are dominant.
Unfortunately, the updated analysis of the $e^+e^-\to\pi^0\pi^0\gamma$ reaction at
$\sqrt{s}\simeq M_\phi^2$ performed by the KLOE Coll.~\cite{Ambrosino:2006hb} with a statistics about thirty times larger does not include an experimental $\pi^0\pi^0$ mass distribution to further test the accuracy of the present approach\footnote{
A detailed explanation of the incorrect treatment of experimental data done in Ref.~\cite{Aloisio:2002bt} is given in Ref.~\cite{Palomar:2003rb}.}.
Anyway, a best fit to $\phi\to\pi^0\pi^0\gamma$ data from Ref.~\cite{Aloisio:2002bt}
was found for $m_{f_0}=985$ MeV and $\phi_S=-8^\circ$ \cite{Escribano:2006mb}.
The value of the $f_0$ mass is in agreement with the recent estimates
$m_{f_0}=976.8^{+10.1}_{-0.7}$ MeV (in the kaon-loop model) by the
KLOE Coll.~\cite{Ambrosino:2006hb} and
$m_{f_0}=985.6^{+1.6}_{-2.2}$ MeV from the study of the $f_0$ resonance in
$\gamma\gamma\to\pi^+\pi^-$ production by the Belle Coll.~\cite{Mori:2006jj}.
The scalar mixing angle used also agrees with the L$\sigma$M prediction
$\phi_S=(-9.1\pm 0.5)^\circ$ \cite{Rodriguez:2004tn}.
As seen in Table \ref{table2}, our former results are in accord with the predicted values given in
Ref.~\cite{Kalashnikova:2005zz},
where the kaon loop mechanism is shown to dominate the $\phi\to f_0\gamma,a_0\gamma$ transitions,
and Ref.~\cite{Ivashyn:2007yy}, where they are computed in a ChPT-inspired approach.
However, they disagree with the predictions given in Ref.~\cite{Black:2002ek}
based on flavour symmetry and vector-meson dominance.
In our approach and that of Refs.~\cite{Kalashnikova:2005zz,Ivashyn:2007yy}
the $V\to S\gamma$ decays proceed through pseudoscalar meson loops,
whereas the production mechanism in Ref.~\cite{Black:2002ek} is driven by intermediate vector mesons,
that is $V\to SV\to S\gamma$, with a pointlike $SVV$ coupling.

Concerning the processes with a $\sigma$ meson in the final state,
our result for the $\rho\to\sigma\gamma$ branching ratio is in agreement with the measured value and the second model (M2) prediction of Ref.~\cite{Black:2002ek}.
In contrast, the difference for the $\omega\to\sigma\gamma$ case is huge.
The reason for such a discrepancy is intrinsic to the approach taken.
While in our case only kaons can propagate in the loop because of isospin conservation,
in Ref.~\cite{Black:2002ek}, not having loops, one would expect a result of the same order of
$\rho\to\sigma\gamma$.
One additional reason for the smallness of our $\omega\to\sigma\gamma$ prediction
is the already mentioned suppression of the $\sigma K\bar K$ coupling in the L$\sigma$M.
The same behaviour occurs for $\phi\to\sigma\gamma$.
Contrary to the $\omega,\phi\to\sigma\gamma$ cases,
pion loops are also allowed for the $\rho\to\sigma\gamma$ reaction and are indeed dominant.
Furthermore, the coupling $\sigma\pi\pi$ is not suppressed.
Notice that for processes involving a $\sigma$ meson the main source of error is by far the uncertainty about the $\sigma$ mass.
In any case, $V\to\sigma\gamma$ transitions merit some caution as advanced in
Ref.~\cite{Black:2002ek}.
Because the $\sigma$ is so broad, the simple two body final state approximation in decays such as
$\rho,\omega,\phi\to\pi^0\pi^0\gamma$ is not accurate and it is better to consider these decays as having three body final states.

\begin{table}
\centerline{
\begin{tabular}{lcccccc}
\hline\\[-2ex]
Observable &
L$\sigma$M &
M1 \cite{Black:2002ek} &
M2 \cite{Black:2002ek} &
\cite{Kalashnikova:2005zz} &
\cite{Ivashyn:2007yy} &
\cite{Nagahiro:2008mn}\\[0.5ex]
\hline\\[-2ex]
$\Gamma_{f_0\to\rho\gamma}$ [keV] &
$24.1^{+21.5}_{-11.6}$ &
$19\pm 5$ &
$3.3\pm 2.0$ &
$3.4$ &
$9.6$ &
$4.2\pm 1.1$\\[0.5ex]
$\Gamma_{f_0\to\omega\gamma}$ [keV] &
$14.1^{+17.1}_{-7.8}$ &
$126\pm 20$ &
$88\pm 17$ &
$3.4$ &
$15.0$ &
$4.3\pm 1.3$\\[0.5ex]
$\Gamma_{a_0^0\to\rho^0\gamma}$ [keV] &
$12.4^{+5.5}_{-0.8}$ &
$3.0\pm 1.0$ &
$3.0\pm 1.0$ &
$3.4$ &
$9.1$ &
$11\pm 4$\\[0.5ex]
$\Gamma_{a_0\to\omega\gamma}$ [keV] &
$11.6^{+5.2}_{-0.8}$ &
$641\pm 87$ &
$641\pm 87$ &
$3.4$&
$8.7$ &
$31\pm 13$\\[0.5ex]
\hline
\end{tabular}
}
\caption{
Comparison of the L$\sigma$M predictions for $S\to V\gamma$ decays with other approaches.}
\label{table3}
\end{table}
With respect to $S\to V\gamma$ decays,
several predictions have appeared  recently in the literature
\cite{Black:2002ek,Kalashnikova:2005zz,Ivashyn:2007yy,Nagahiro:2008mn}.
Though there are not yet measurements to compare with, these processes are very interesting since they can also shed light on the properties and structure of the lightest scalar mesons and hence complement the information obtained from $\phi\to f_0\gamma,a_0\gamma$ decays.
The result of a first test measurement of $a_0,f_0\to V\gamma$ decays at COSY,
based on $pd\to{^3}\mbox{He}\,a^0_0/f_0$ and $pd\to t a_0^+$ data taken at WASA in November 2007,
is expected to come soon \cite{Buscher:2008ge}.
Our predictions for the $f_0,a_0\to\rho\gamma,\omega\gamma$ processes are shown in
Table \ref{table3}.
The large errors reflect the uncertainties about the scalar masses and the choice of kaon threshold.
This time, however, the most important error comes from the uncertainty about the $f_0$ mass for $f_0$ initiated decays and the choice of kaon threshold for $a_0$ decays.
As seen, some predictions can even differ in orders of magnitude when compared to others, showing the importance of measuring these decays in order to accept or reject one model or the other.
In particular, our results reasonably agree with those of Ref.~\cite{Ivashyn:2007yy}.
This can be understood by the fact that both approaches are inspired in chiral symmetry and use meson loops as the production mechanism.
However, in the ChPT-based approach used in Ref.~\cite{Ivashyn:2007yy}
the $SPP$ couplings $c_{d,m}$ are fixed from phenomenology,
whereas in the L$\sigma$M these couplings are predicted to be $|c_d|=2|c_m|=f_\pi/\sqrt{2}$
\cite{Bramon:2003xq}.
For the $a_0^0\to\rho^0\gamma$ process, we also agree with Ref.~\cite{Nagahiro:2008mn}
where these decays are evaluated in a chiral unitary approach and the $a_0$ and $f_0$ scalar mesons are dynamically generated.
One interesting feature of the approach in Ref.~\cite{Nagahiro:2008mn}
is the computation of vector meson loop effects which do not seem to play a relevant role apart from the $a_0^0\to\omega\gamma$ case\footnote{
Our prediction for this process would agree with their result if the vector meson loop contribution were not considered.}.
However, this new type of contribution driven by a loop made of a vector and a pair of pseudoscalar mesons which subsequently rescatter into a scalar meson may be relevant not only in some decays involving the $\omega$ but also in processes such as
$\phi\to\sigma\gamma$ and $\omega\to\sigma\gamma$ (in our approach dominated by kappa loops) where the kaon loops are not the main contribution.
For instance, in Ref.~\cite{Palomar:2003rb} it is shown that the contribution of vector meson loops to the $\pi^0\pi^0$ invariant mass spectrum of the $\phi\to\pi^0\pi^0\gamma$ decay is similar in magnitude to that of chiral loops in the region around 500 MeV where the resonant effects of the $\sigma$ meson should be visible.
Finally, in Ref.~\cite{Kalashnikova:2005zz} a single scalar coupling $g_S^2/(4\pi)=0.6\ \mbox{GeV}^2$
is used to obtain $\Gamma(f_0/a_0\to\gamma\rho/\omega)=3.4$ keV in the kaon loop model.
In our framework, the $SPP$ couplings are fixed in terms of pseudoscalar and scalar masses,
decay constants and the scalar mixing angle ---see Eq.~(\ref{ASPP}).
Numerically, the required couplings are found to be
$g_{f_0 K\bar K}^2/(4\pi)=1.4\ \mbox{GeV}^2$ and $g_{a_0 K\bar K}^2/(4\pi)=0.9\ \mbox{GeV}^2$,
thus giving larger decay widths.
Notwithstanding, the authors of Ref.~\cite{Kalashnikova:2005zz} argue that kaon and quark loops yield 
contributions of the same order to these processes.
Obviously, quark loops are not incorporated in our approach.

\section{Conclusions}
\label{conclusions}
In this Letter we have worked out the Linear Sigma Model (L$\sigma$M) predictions of radiative
$V\to S\gamma$ and $S\to V\gamma$ decays with $V=\rho,\omega,\phi$ and $S=a_0,\sigma,f_0$.
The contributions arising from kaon, pion and kappa loops have been computed and their relative weights for each process evaluated.
We have seen that kaon loop effects are dominant in all processes besides $\rho\to\sigma\gamma$ and $\omega,\phi\to\sigma\gamma$, where pion and kappa loops are the main contribution, respectively.
Our results are affected by large errors due to uncertainties about the scalar masses and the choice of kaon threshold.
In particular, we have achieved a reasonable agreement with current experimental values for
$\phi\to f_0\gamma$ and $\rho\to\sigma\gamma$ branching ratios and the ratio
$R_{\phi\to f_0\gamma/a_0\gamma}$.
This is not the case of the $\phi\to a_0\gamma$ branching ratio whose measured value is in conflict with all approaches including ours.
We are confident about the validity of our former results since they have been somewhat investigated in
$V\to P^0P^0\gamma$ decays where scalar effects are the main contribution.
Concerning $f_0,a_0\to\rho\gamma,\omega\gamma$ decays, our predictions are quite in accord with other approaches based on chiral symmetry and pseudoscalar meson loops.
To conclude, the L$\sigma$M has been shown to very useful for extracting relevant information on the properties and nature of the lightest scalar mesons from the analysis of radiative
$V\to S\gamma$ and $S\to V\gamma$ decays.
More precise measurements by the KLOE Coll., the ongoing experiment at WASA@COSY, and the future facilities DAFNE-2 (Frascati) and VEPP-2000 (Novosibirsk) will make possible to test the goodness of the whole approach.

\section*{Acknowledgements}
This work was supported in part by the Ramon y Cajal program,
the Ministerio de Educaci\'on y Ciencia under grant FPA2005-02211,
the EU Contract No.~MRTN-CT-2006-035482, ``FLAVIAnet'',
the Spanish Consolider-Ingenio 2010 Programme CPAN (CSD2007-00042), and
the Generalitat de Catalunya under grant 2005-SGR-00994.

\end{document}